\documentclass[conference]{IEEEtran}
\IEEEoverridecommandlockouts
\usepackage{cite}
\usepackage{amsmath,amssymb,amsfonts}
\usepackage{algorithmic}
\usepackage{graphicx}
\usepackage{subcaption}
\usepackage{textcomp}
\usepackage{xcolor}
\usepackage{listings} 
\usepackage{braket}  
\usepackage{quantikz}
\usepackage{multirow}
\usepackage{tikz}
\usepackage[skip=0pt]{caption}
\usepackage{hyperref} 

\def\BibTeX{{\rm B\kern-.05em{\sc i\kern-.025em b}\kern-.08em
    T\kern-.1667em\lower.7ex\hbox{E}\kern-.125emX}}

\begin{document}

\title{Benchmarking Quantum Architecture Search with Surrogate Assistance\\
\thanks{This work was partially funded by the BMWK project EniQmA (01MQ22007A). Johannes Jung and Daniel Barta contributed equally to this work.}
}
\author{\IEEEauthorblockN{ Darya Martyniuk\textsuperscript{1},  Johannes Jung\textsuperscript{1, *},  Daniel Barta\textsuperscript{1, *},  Adrian Paschke\textsuperscript{1, 2}}
\IEEEauthorblockA{
  \textsuperscript{1 }Fraunhofer FOKUS, Berlin, Germany\\
        \textsuperscript{2} Freie Universität Berlin, Berlin, Germany\\
        Email: \{darya.martyniuk, johannes.jung, daniel.barta\}@fokus.fraunhofer.de
}
}
\maketitle

\begin{abstract}
The development of quantum algorithms and their practical applications currently relies heavily on the efficient design, compilation, and optimization of quantum circuits.
In particular, parametrized quantum circuits~(PQCs), which serve as the basis for variational quantum algorithms~(VQAs), demand carefully engineered architectures that balance performance with hardware constraints.
Despite recent progress, identifying structural features of PQCs that enhance trainability, noise resilience, and overall algorithmic performance remains an active area of research. 
Addressing these challenges, quantum architecture search~(QAS) aims to automate the design of problem-specific PQCs by systematically exploring circuit architectures to optimize algorithmic performance, often with varying degrees of consideration for hardware constraints.
However, comparing QAS methods is challenging due to the absence of a unified benchmark evaluation pipeline, and the high resource demands. 
In this paper, we present SQuASH, the Surrogate Quantum Architecture Search Helper, a benchmark that leverages surrogate models to enable uniform comparison of QAS methods and considerably accelerate their evaluation.
We present the methodology for creating a surrogate benchmark for QAS and demonstrate its capability to accelerate the execution and comparison of QAS methods. 
Additionally, we provide the code required to integrate SQuASH into custom QAS methods, enabling not only benchmarking but also the use of surrogate models for rapid prototyping. 
We further release the dataset used to train the surrogate models, facilitating reproducibility and further research.
\end{abstract}


\section{Introduction}
Advances in artificial intelligence (AI) research over the past decade
motivated the exploration of AI-driven techniques to address the challenges in QC, including compilation~\cite{tang2024alpharouter}, device control~\cite{wright2023fast, vora2024ml}, error correction~\cite{wang2023transformer, chamberland2023techniques} and mitigation~\cite{liao2024machine}, automated quantum circuit design and optimization~\cite{paradis2024synthetiq, nakaji2024generative, patel2024curriculum, bilkis2021semi,furrutter2024quantum, li2024quarl}. 
Quantum architecture search~(QAS)~\cite{zhang2022differentiable, martyniuk2024quantum} is a subfield of the automated quantum circuit design research, focusing on generation and optimization of parametrized quantum circuits~(PQCs). 
Despite the potential of QAS to improve the efficiency and performance of quantum circuits and discover novel architectures, benchmarking of QAS approaches is challenging due to the significant computational resources required for empirical evaluation. 
This is mainly because most QAS methods involve two key components: The training of a search algorithm to identify the optimal circuit for a given task and the training of the candidate PQCs that the search algorithm explores during the search process. 
Furthermore, the evaluation of QAS methods currently relies on custom-designed search spaces, which significantly hinders reproducibility and fair comparison across approaches.
A standardized and accessible benchmark would provide common ground for evaluations.
Inspired by the concept of surrogate benchmarks in the classical neural architecture serch~(NAS) field,  which aim to automate the design of optimal neural network architectures, we introduce SQuASH, the \underline{S}urrogate \underline{Q}uantum \underline{A}rchitecture \underline{S}earch \underline{H}elper. 
SQuASH is designed as a QAS benchmark, covering three predefined search spaces for two tasks: quantum state preparation and linear classification. 
In addition, for each search space, the benchmark provides pre-trained surrogate models, which are classical regression models designed to predict the performance of PQCs after optimization. 

We summarize our \textbf{contributions} as follows: (1)~We present an extendable surrogate QAS benchmark that enables and significantly accelerates QAS methods comparison and prototyping, providing evaluation protocols and an easy integration into custom approaches. 
(2)~We evaluate the ability of surrogate models to predict the performance of PQCs on a given task after training.
(3)~To encourage the use, reproducibility and further development of the benchmark, we open source\footnote{\url{
https://github.com/SQuASH-bench/SQuASH}} the data and code for surrogate models. 

This work is structured as follows. 
Section~\ref{sec:background} reviews related work. 
Section~\ref{sec:methods} describes the methodology used to create the benchmark 
and outlines the 
data collection and training process for surrogate models.
Section~\ref{sec:evaluation} presents evaluation results, followed by a discussion on potentials and limitations in Section~\ref{sec:limitations}. Section~\ref{sec:conclusion} summarizes the main findings and concludes the paper.

\section{Related Work}\label{sec:background}

\subsection{Quantum Architecture Search (QAS)}
Automated quantum circuit synthesis and optimization apply heuristic and machine learning methods to design and refine quantum circuits across different layers of abstraction. 
The target problems range from specific sub-tasks, such as qubit mapping,
routing
or minimizing the number of specific gate types
\cite{paler2023machine, li2024pauliforest, pozzi2022using, LeCompte, tang2024alpharouter}, to the task-specific design of quantum circuits with explicit consideration of noise~\cite{patel2024curriculum, wang2022quantumnas, anagolum2024elivagar}.
To achieve this, different paradigms are employed, including reinforcement learning~\cite{Altmann, pozzi2022using,tang2024alpharouter,patel2024curriculum}, genetic algorithms~\cite{chivilikhin2020mog, sunkel2023ga4qco}, adaptive strategies~\cite{grimsley2019adaptive, bilkis2021semi}, Bayesian optimization~\cite{duong2022quantum}, or generative models~\cite{nakaji2024generative, Frrutter2023QuantumCS}. 
QAS refers to approaches designed to automate the search for an optimal PQC with respect to its structure $U(\overrightarrow{\theta})$ and trainable parameters~$\overrightarrow{\theta}$. 
Since PQCs share training similarities with classical neural networks - and are sometimes referred to as quantum neural networks in the context of quantum machine learning~(QML) - QAS methods often benefit from advances in classical NAS. Yet, these advances must be carefully adapted to align with the constraints and operational principles of quantum systems.
For a comprehensive overview of quantum circuit synthesis and QAS, we refer to surveys~\cite{ge2024quantum, martyniuk2024quantum}.

QAS approaches vary in their strategies for addressing parameter optimization. 
Certain techniques~\cite{wang2022quantumnas,sun2023differentiable, patel2024curriculum, bilkis2021semi, duong2022quantum} treat parameter optimization as an inner loop within the architecture search process: The algorithm first proposes a circuit structure, which is then trained using a classical optimizer such as COBYLA~\cite{Powell1994ADS} or Adam~\cite{kingma2014adam}. 
Techniques like weight-sharing~\cite{du2022quantum, wang2022quantumnas} can be employed at this stage to improve training efficiency.
In contrast, other methods~\cite{nakaji2024generative, Altmann} integrate parameter optimization directly into the search procedure itself, aiming to simultaneously generate circuit architectures with their optimal parameters. 
A third category of approaches leverages surrogate models~\cite{Zhang2021NeuralPB, he2023gnn, he2023gsqas} or training-free performance predictors~\cite{anagolum2024elivagar, ZhiminTrainingFree} to approximate the outcome of PQC training, i.e., its performance, thus reducing or bypassing the need for full training of candidate circuits.
The \textit{benchmark} introduced in this paper builds upon the idea of using classical models to approximate the performance of PQCs after training. 
In contrast to previous work, we explicitly define the search space, and release not only the code for the surrogate models, but also multiple pretrained models and the underlying training data, thereby offering a reproducible and extensible benchmark for future work. 
While the benchmark tasks can be used to evaluate all QAS approaches, the surrogate-based evaluation is particularly suited for methods that treat parameter optimization as a separate inner loop, since they benefit from skipping full circuit training. 
For this reason, our surrogate-based benchmark setup is aligned with this class of methods and assumes a basic variant without weight-sharing.
However, the \textit{dataset} provided with the benchmark includes both initial and optimized circuits, making it useful for training of QAS methods that integrate parameter optimization directly into the search process.

\subsection{Benchmarks and Datasets for QAS }
With the development of large high-quality cross-benchmarking efforts in quantum computing, e.g., QData~\cite{perrier2022qdataset}, Hamlib~\cite{sawaya2024hamlib}, or MQTBench~\cite{quetschlich2023mqtbench}, the landscape of benchmarks specifically tailored to QAS methods remains notably limited.
QAS-Bench~\cite{QASBench} fills this gap by evaluating automated circuit design techniques through two tasks: quantum circuit regeneration and unitary approximation. 
For each task, the benchmark offers a dataset comprising 900 circuits for quantum circuit regeneration and 400 unitary matrices along with a defined evaluation protocol and six baseline methods. 
In parallel, several environments have been introduced to support reinforcement learning~(RL) agents in the context of QAS. Ref.~\cite{altmann2023challenges} presents \texttt{gcd-gym}, a quantum circuit design environment designed for generic state preparation and unitary composition tasks.  
Similarly,~\cite{van2023qgym} introduce~\texttt{qgym}, a customizable RL environment for quantum compilation. 
While prior contributions tailored for QAS benchmarking provide useful tools, they do not address the high computational costs associated with comparing QAS methods and often focus on specific learning paradigms, e.g., RL. 
Our benchmark addresses these limitations by supporting various algorithmic strategies, standardizing the search space and task definitions, and accelerating experimentation through efficient surrogate-based evaluation.
It is highly inspired by the success of surrogate-based NAS benchmarks~\cite{zela2020surrogate} in the classical community, which address the high computational demands of NAS. 

In addition to benchmarking tools, data is essential in enabling ML-based QAS methods. 
It can be used to train surrogate models for efficient performance estimation, provide demonstrations that enhance model generalization, or function as complete training datasets, especially for generative modeling techniques. 
QCircuitNet~\cite{yang2024qcircuitnet} is a structured dataset for quantum algorithm design at the code level, ranging from basic primitives to advanced applications. Each instance includes problem descriptions, circuit generation code, algorithm circuits, oracles, and post-processing functions, designed to support LLM-driven algorithm synthesis. 
Ref.~\cite{Frrutter2023QuantumCS} introduces the dataset prepared to train a diffusion model for the entanglement generation and unitary compilation tasks. Similarly, the NTangled dataset~\cite{schatzki2021entangled} comprises quantum states with varying degrees of entanglement. 
Motivated by the task of classifying and clustering quantum circuits based on the similarity of output states,~\cite{nakayama2023vqe} introduces a dataset consisting of PQCs optimized by the Variational Quantum Eigensolver~(VQE). 
This dataset includes six common types of condensed matter Hamiltonians over different systems and circuit structures. 
To support ongoing initiatives in data sharing within quantum computing research and facilitate community-driven extensions of the benchmark, we contribute a large-scale dataset aligned with the defined benchmark search spaces.
It contains more than 3M unique PQCs in total, along with information on the circuit parameters both before and after optimization.

\section{Benchmark design}\label{sec:methods}
\begin{figure}[tb]
\centerline{\includegraphics[width=0.95\linewidth]{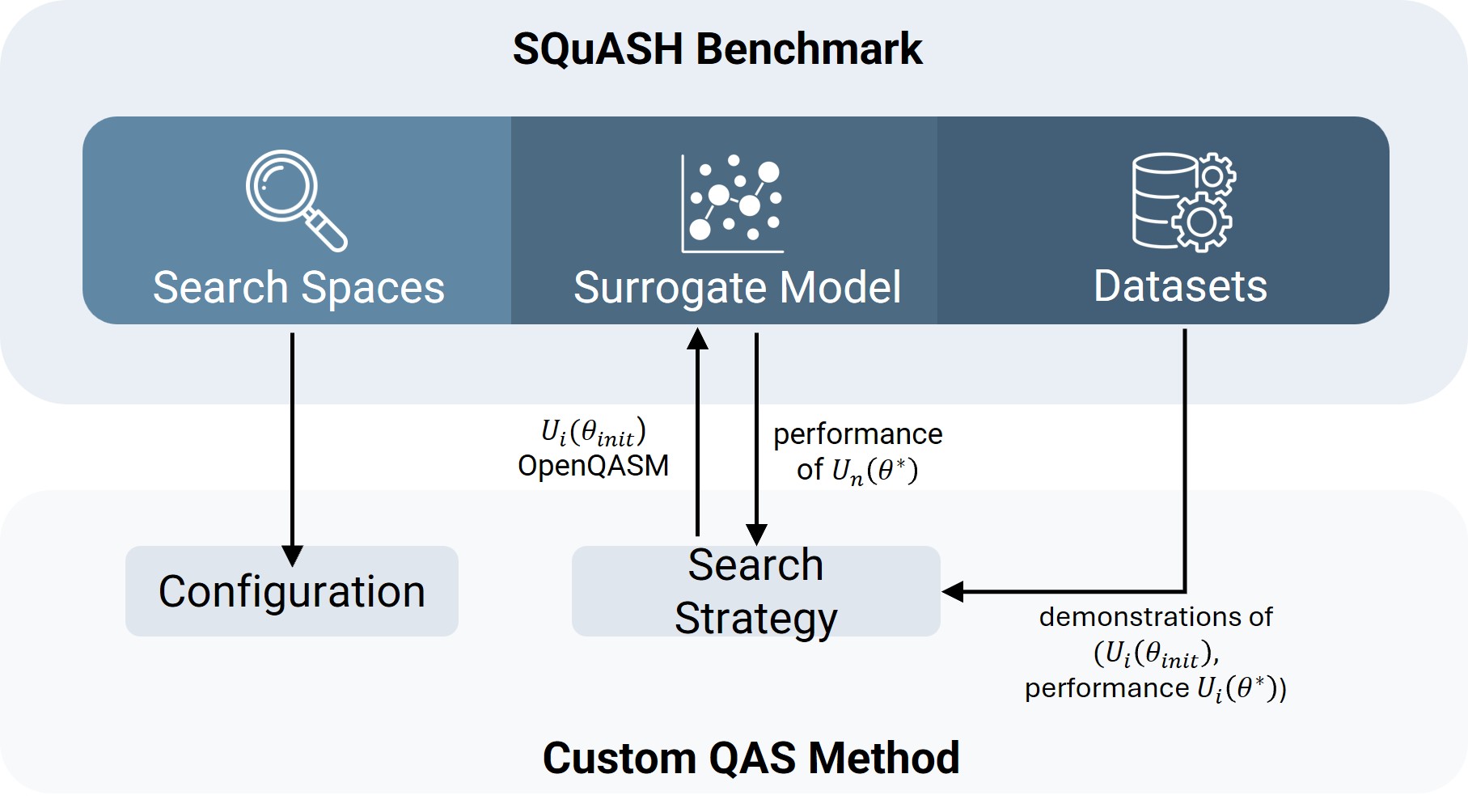}}
\caption{Integration of the SQuASH benchmark and its artifacts into a custom QAS pipeline.}

\label{fig:qas_usage}
\end{figure}

The goal of the SQuASH benchmark~\footnote{\url{https://squash-benchmark.fokus.fraunhofer.de/}} is two-fold. First, it provides researchers with \textit{different design tasks} 
and the corresponding \textit{evaluation protocols}, enabling a reproducible comparison of the QAS methods, independent of the search strategy used. 
Second, 
the benchmark provides pre-trained \textit{surrogate models} for selected search spaces.
A surrogate model, implemented as a classical regression model, takes an untrained candidate PQC generated during the circuit design process and predicts its expected performance after training. 
This means, instead of fully training the candidate circuits and subsequently evaluating them on quantum hardware, the QAS algorithm can query the surrogate model to predict the final performance of the PQC, as depicted in Fig.~\ref{fig:qas_usage}. 
This allows researchers to drastically reduce the runtime of QAS methods during the development phase.
Although predictions will not be exact, \cite{zela2020surrogate} shows that for classical neural network architectures they can be highly accurate, which is also to be expected for PQC architectures.
Moreover, if the search strategy benefits from demonstrations or additional learning, as in the case of off-policy reinforcement learning, the demonstration buffer can be populated by querying the datasets. 
\subsection{Search Spaces}
For the first version of the benchmark, we consider two distinct tasks: quantum state preparation and data point classification. For each task, we define its objective, describe the available search spaces, and outline the evaluation pipelines used to evaluate QAS methods.
\subsubsection{Quantum state preparation}
Preparing quantum states efficiently is a fundamental subroutine in quantum computing. A state preparation procedure constructs a quantum circuit that transforms an initial $n$-qubit state
$\ket{0}^{\otimes n} = \ket{00\cdots0}$ into a desired target state~$\ket{\psi} \in \mathbb{C}^{2^{n}}$. For this benchmark, we assume that QAS methods need to prepare the 3-qubit maximally entangled state, the Greenberger-Horne-Zeilinger~(GHZ) state:
$\ket{\text{GHZ}_3} = \frac{1}{\sqrt{2}} \left( \ket{000} + \ket{111} \right)$. 
To ensure comparability across different QAS methods and improve the reliability of surrogate model predictions, we constrain the search space by explicitly specifying the configurations that the QAS methods are allowed to use when solving the task. 

The search space is assumed to be gate-wise, meaning that at each step, a QAS algorithm is allowed to add or remove a single gate (rather than a layer or block of gates) from the circuit. 
If candidate circuits contain parameterized gates, the initial parameters are assumed to be drawn from a uniform distribution within a small range around $0.0$, specifically $[-0.01, 0.01]$, and are subsequently optimized using the COBYLA optimizer for a maximum of 300 steps. 
The optimization bounds are set to $[-\pi, \pi]$. 
The benchmark specifies two allowed gate sets: \textit{gs1} = \{'cx', 'h', 'rx', 'ry', 'rz', 'id'\} and \textit{gs2} = \{ 'cz', 'id', 'rx', 'rz', 'rzz', 'sx', 'x'\}, whereby the \textit{gs2} represents the native gate set of the \textit{ibm\_fez}, which is one of the IBM quantum processors.
In both cases, the number of qubits is limited to 3, and the maximal depth of the non-transpiled circuit is restricted to 6 for \textit{gs1} and 10 for \textit{gs2}, respectively. 
The quantum environment is assumed to be noise-free and the circuits are executed on the Qiskit~\cite{qiskit2024} quantum simulator~\textit{'statevector\_simulator'} with 1000 shots. 
Table~\ref{tab:ghz-search-space} summarizes two predefined search spaces for this task.

\begin{table}[htbp]
\caption{Benchmark Search Spaces for State Preparation}
\begin{center}
\begin{tabular}{|c|cc|}
\hline
                                                                        & \multicolumn{1}{c|}{\textbf{GHZ\_a}}                                                           & \textbf{GHZ\_b}                                                                                                                    \\ \hline
Target State                                                            & \multicolumn{2}{c|}{$\ket{\text{GHZ}_3} = \frac{1}{\sqrt{2}} \left( \ket{000} + \ket{111} \right)$}                                                                                                                                                 \\ \hline
Search Space Type                                                       & \multicolumn{2}{c|}{Gate-wise}                                                                                                                                                                                                                      \\ \hline
Number of qubits                                                        & \multicolumn{2}{c|}{3}                                                                                                                                                                                                                              \\ \hline
Gate Set                                                                & \multicolumn{1}{c|}{\begin{tabular}[c]{@{}c@{}}\{'cx', 'h', 'rx',\\ 'ry', 'rz', 'id‘\}\end{tabular}} & \begin{tabular}[c]{@{}c@{}}\{'cz', 'id', 'rx', \\ 'rz', 'rzz', 'sx', 'x‘\}\\ (native of \textit{ibm\_fez})\end{tabular} \\ \hline
Initial Parameters                                                      & \multicolumn{1}{c|}{{[}-0.01, 0.01{]}}                                                                   & {[}-0.01, 0.01{]}                                                                                                                            \\ \hline
Max. Depth                                                              & \multicolumn{1}{c|}{6}                                                                                 & 10                                                                                                                                         \\ \hline
Optimizer                                                               & \multicolumn{2}{c|}{COBYLA}                                                                                                                                                                                                                         \\ \hline
\begin{tabular}[c]{@{}c@{}}Max. Optimization \\ Iterations\end{tabular} & \multicolumn{2}{c|}{300}                                                                                                                                                                                                                            \\ \hline
Optimization Bounds                                                     & \multicolumn{2}{c|}{[-$\pi$; $\pi$]}                                                                                                                                                                                                                  \\ \hline
Quantum Backend                                                         & \multicolumn{2}{c|}{\textit{'statevector\_simulator'}}                                                                                                                                                                                              \\ \hline
\end{tabular}
\label{tab:ghz-search-space}
\end{center}
\end{table}
\begin{table}[htbp]
\begin{center}
\caption{Benchmark Search Spaces for Classification}
\begin{tabular}{|c|c|}
\hline
\textbf{}                                                               & \textbf{LS\_a}                                                                             \\ \hline
Task                                                       & Linear classification    \\ \hline 
Search Space Type                                                       & Gate- and Layerwise                                                                               \\ \hline
\multirow{2}{*}{Gate Set}                                               & Layers: \{RxRotation, RyRotation\}                                                                \\ \cline{2-2} 
                                                                        & \begin{tabular}[c]{@{}c@{}}Gates: \{'rx', 'ry', 'cx', \\ 'h', 'swap', 'crx', 'cry'\}\end{tabular} \\ \hline
Dataset                                                                 &              \begin{tabular}[c]{@{}c@{}}Linearly Separable \\ (300 samples, n\_features=8, margin=1)\end{tabular}                            \\ \hline
Data Encoding                                                           & \begin{tabular}[c]{@{}c@{}}Angle Encoding \\ (2 features per qubit via $rx$, $ry$)\end{tabular}          \\ \hline
No. of Qubits                                                        & 4                                                                                                 \\ \hline
Initial Parameters                                                      & {[}-0.01, 0.01{]}                                                                                   \\ \hline
Max. Ansatz Depth                                                              & 10                                                                                                \\ \hline
Optimizer                                                               & COBYLA                                                                                            \\ \hline
\begin{tabular}[c]{@{}c@{}}Max. Optimization \\ Iterations\end{tabular} & 200                                                                                               \\ \hline
Quantum Backend                                                         &  {\textit{'statevector\_simulator'}}                                                                                                 \\ \hline
\end{tabular}
\label{tab:ls-search-space}
\end{center}
\end{table}

\subsubsection{Data Points Classification}
The fundamental question of what constitutes an effective, scalable and a high-performing QML model remains under active investigation.
From our perspective, benchmarking QAS methods for QML tasks should align with benchmark datasets used to evaluate QML models with fixed ansätze, enabling comparisons not only between different QAS methods but also with manually designed models.
Recently,~\cite{bowles2024better} introduced a large-scale benchmark study aimed at systematically evaluating popular QML models across different classification tasks. 
In this work, we reuse the \texttt{LinearlySeparable} dataset from~\cite{bowles2024better}, which corresponds to a linear data points classification task. 
To produce the dataset, we follow the data generation procedure outlined in the original work and set the random seed to 274, number of features~$d = 8$, number of samples~$N = 300$, perceptron weights~$w = (1, ..., 1)^T$ and margin~$m = 1$. 
After the split, the training dataset consists of 240 points, and the test dataset contains 60 points.

In QAS methods, each candidate PQC in this search space, consists of a \textit{fixed feature encoding block} and a \textit{generated ansatz}. We use angle encoding with two rotation gates,  where each qubit encodes two features using sequential $rx$ and $ry$ rotations. As a result, the candidate circuits have a width of~4~qubits.
The circuit ansatz is limited to a maximum depth of~10 before transpilation. 
Since QML models rely heavily on rotations, rotation layers composed of $rx$ or $ry$ gates are allowed in this search space for the ansatz generation, in addition to the gates from the specified gate set  \textit{gs3} = \{’rx’, ’ry’, ’cx’, ’h’, ’swap’, ’crx’, ’cry’\}.
Table~\ref{tab:ls-search-space} summarizes the definition of the search space, which we refer to as \textit{LS\_a} throughout the remainder of the paper.
\subsection{Data Collection}
We constructed the training dataset for the surrogate models by exploring the predefined search spaces using three strategies: reinforcement learning (RL), a genetic algorithm (GA), and random search. 
Each strategy was executed with multiple random seeds.
In the following, we provide details on the data collection procedures used for each task.
\subsubsection{Quantum State Preparation}\label{sec:data_collection_quant_state_prep}
Each search strategy generates and optimizes candidate PQCs with the objective of approximating the target quantum state $\ket{\text{GHZ}_3}$. The primary evaluation criteria guiding the process of parameter optimization in candidate circuits is the fidelity between the $\ket{\text{GHZ}_3}$ and the state produced by the candidate circuit. A fidelity value of 1 indicates an exact match and 0 reflects maximal divergence.
For every circuit produced, we save the circuit and its associated metadata in a serialized \texttt{.pickle} file. 
Specifically, each file contains the following information: (i)~the initial PQC as a QASM3 string, representing the circuit with parameter values prior to optimization; (ii)~the optimized PQC as a QASM3 string; (iii)~the fidelity score, measuring the circuit’s accuracy with respect to the target quantum state; and (iv)~a circuit hash, a unique SHA-256 hash of the QASM3 representation of the initial circuit.
Once the data has been collected, we apply a filtering on the entire data to remove duplicate circuits based on their hash values. 
This ensures that only unique circuits (both regarding the structure and parameters) are retained in the dataset. 

\begin{figure}[tb]
\centerline{\includegraphics[width=1\linewidth]{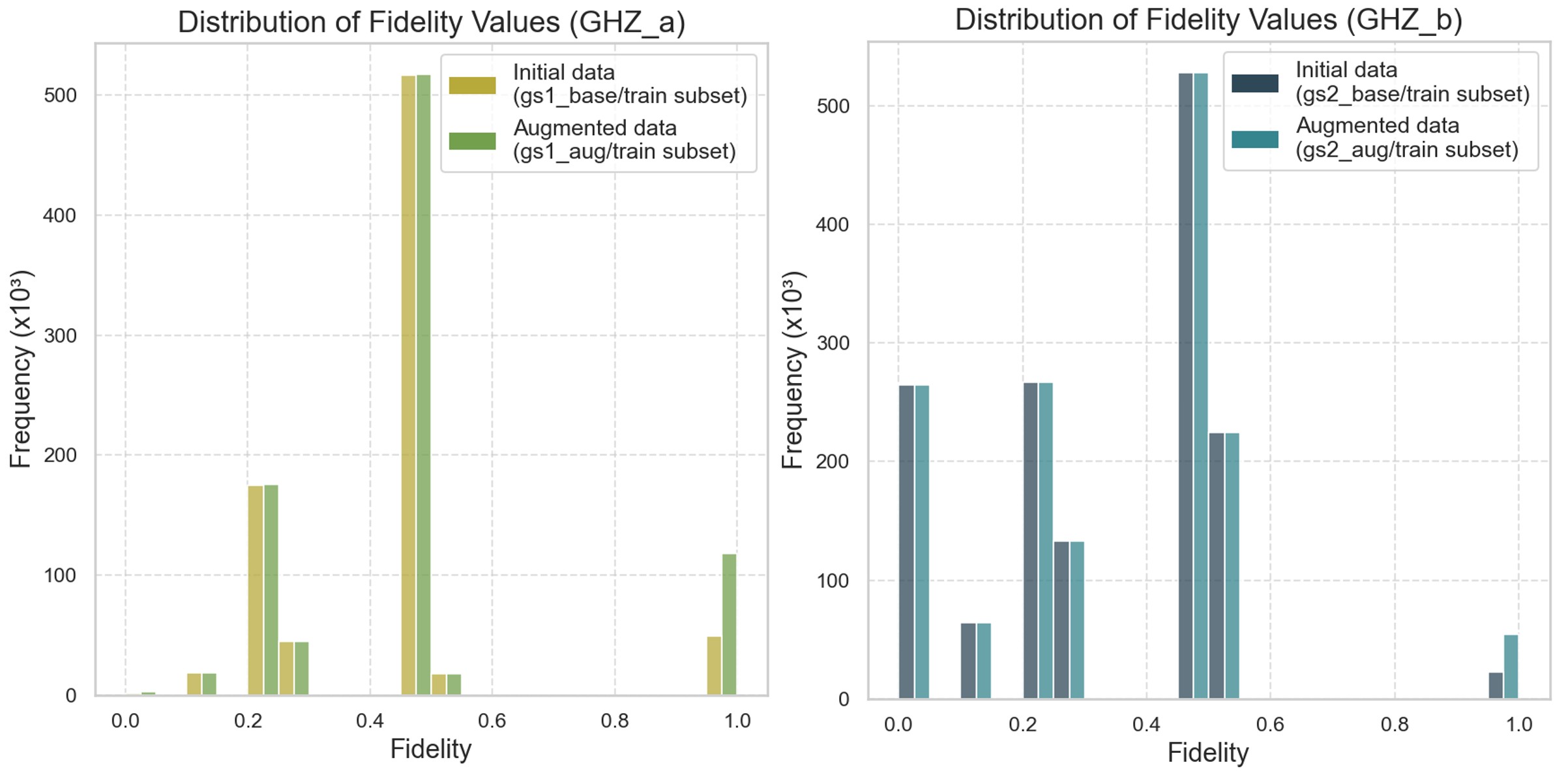}}
\caption{Circuit fidelity distribution in training data for state preparation search spaces.}
\label{fig:circuit_fid}
\end{figure}

Next, we analyzed the fidelity distribution within datasets and identified imbalances for both \textit{GHZ\_a} and \textit{GHZ\_b}.
For \textit{GHZ\_a}, a large number of circuits achieved fidelity around~0.5.
To address this, we performed an additional random search using~10 different seeds and stored all generated circuits except those with fidelity values near~0.5. 
Additionally, to better represent underexplored regions of the search space, specifically, high-fidelity circuits and those with fidelity near~0.0, we performed targeted data augmentation. 
We extracted relevant circuits and transpiled them using a reduced gate set (e.g., by removing \textit{h} or \textit{ry} gates) and different optimization levels. If the resulting fidelity remained at 0.0 or reached~$\geq 0.99$ and its hash was unique, the transpiled circuit was added to the dataset. 
In the following, we refer to the original and augmented versions of the \texttt{GHZ\_a} dataset as \textit{gs1\_base} and \textit{gs1\_aug}, respectively.
For \textit{GHZ\_b},  we denote the collected data as~\textit{gs2\_base}. 
In~\textit{gs2\_base}, only a small subset of circuits demonstrate fidelity values close to~1.0.
 To enhance diversity, we selected all circuits with fidelity equal to~1.0 from~\textit{gs1\_base} and transpiled them using the gate set of \texttt{GHZ\_b}. 
 Only transpiled circuits that conformed to the predefined constraints of the \texttt{GHZ\_b} search space, e.g., maximum depth~$\leq$~10, was retained and incorporated into the dataset. 
Additionally, we applied the same procedure to the augmented high-fidelity circuits from \textit{gs1\_aug}, yielding a second dataset variant, denoted as \textit{gs2\_aug}. 
Fig.~\ref{fig:circuit_fid} shows the distribution of circuits in augmented and non-augmented dataset versions.
\subsubsection{Classification}\label{sec:data_collection_classification}
\begin{figure}[tb]
\centerline{\includegraphics[width=0.95\linewidth]{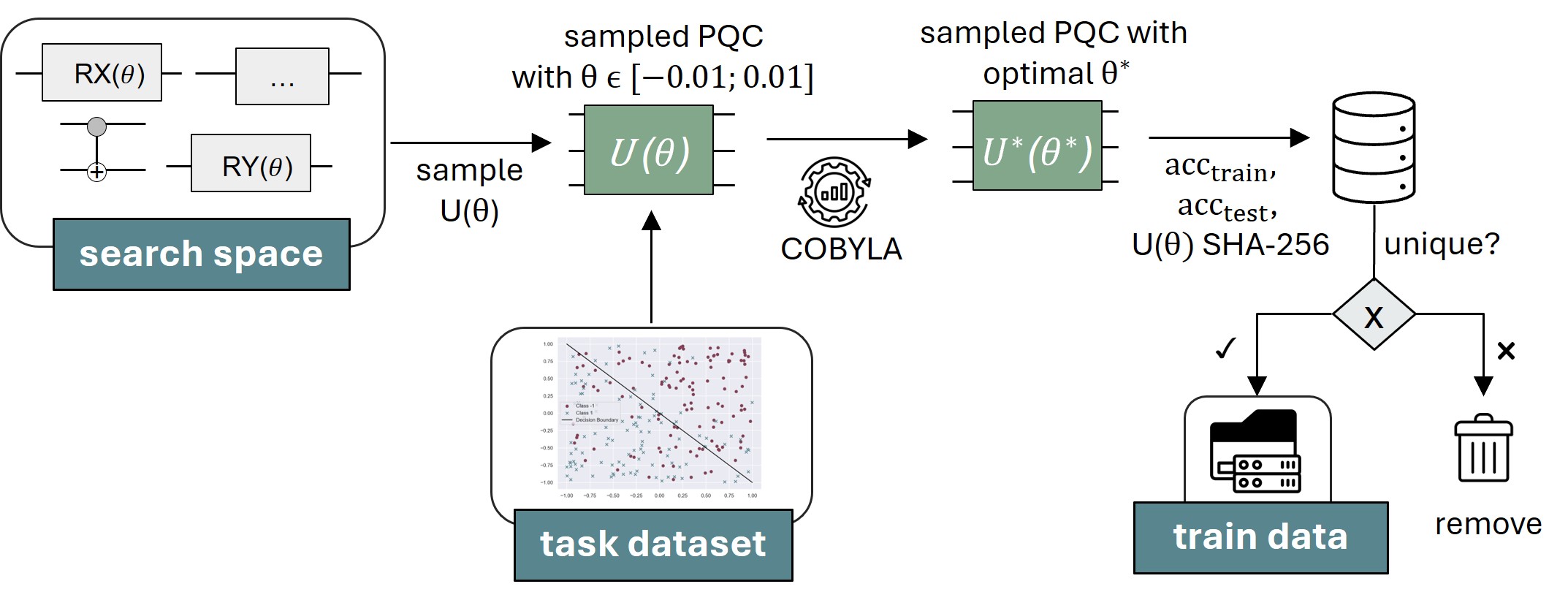}}
\caption{Data collection process for \textit{ls\_base}. A candidate PQC is iteratively constructed by sampling gates or layers from predefined gate and layer sets. Each generated PQC is trained on the full training subset and then evaluated on the test subset. If the initial (i.e., untrained) circuit, along with its parameters, has not been encountered previously, it is stored in the dataset and saved as a persistent file for future use.}
\label{fig:data_collection}
\end{figure}

In this task, similar to the previous one, candidate PQCs are optimized in terms of both structure and parameters. 
However, in contrast to the state preparation task, the goal here is to construct a PQC that effectively performs linear classification. The dataset of PQCs for this search space, denoted in following as \textit{ls\_base}, was collected in a manner similar to the state preparation task, using the same RL, GA, and random search implementations. Fig.~\ref{fig:data_collection} schematically illustrates the data collection process. 
The circuit metadata in \textit{ls\_base} stores the OpenQASM string of (i)~the initial ansatz without encoding, and (ii)~the optimized ansatz. It also includes (iii)~the hash of the initial circuit, (iv)~the train and (v)~test accuracy of the optimized circuit, replacing the previous fidelity metric.
Table~\ref{tab:datasets} presents a summary of the statistics for all dataset versions in SQuASH. 

\renewcommand{\arraystretch}{1}
\begin{table}[tb]
\caption{SQuASH datasets and their sizes}
\begin{tabular}{|@{\hskip 0.2pt}c@{\hskip 0.2pt}|@{\hskip 0.2pt}c@{\hskip 0.2pt}|@{\hskip 0.2pt}c@{\hskip 0.2pt}|@{\hskip 0.2pt}c@{\hskip 0.2pt}|c@{\hskip 0.2pt}c@{\hskip 0.2pt}|@{\hskip 0.2pt}c@{\hskip 0.2pt}|}
\hline
\textbf{Task}         & \multicolumn{4}{c|}{\textbf{State preparation}}                                                                                    & \textbf{Classification} \\ \hline
\textbf{Search space} & \multicolumn{2}{c|}{\textbf{GHZ\_a}}                                       & \multicolumn{2}{c|}{\textbf{GHZ\_b}}                  & \textbf{LS\_a}          \\ \hline
\textbf{Dataset}      & \multicolumn{1}{c|}{gs1\_base} & \multicolumn{1}{c|}{gs1\_aug} & \multicolumn{1}{c|}{gs2\_base} & \multicolumn{1}{c|}{gs2\_aug} & {ls\_base}             \\ \hline
\textbf{Train subset} & \multicolumn{1}{c|}{823.562}      & \multicolumn{1}{c|}{895.271}           & \multicolumn{1}{c|}{1.503.854}    & \multicolumn{1}{c|}{1.535.841}         & {311.108}                 \\ \hline
\textbf{Test subset}  & \multicolumn{2}{c|}{82.926}                                                & \multicolumn{2}{c|}{97.753}                           & 34.569                  \\ \hline
\end{tabular}
\label{tab:datasets}
\end{table}
\subsection{Surrogate Models}\label{sec:surrogate}
In this work, we train the surrogate model to predict a target metric $y_{i}$, e.g., fidelity or training accuracy, based on the given initial PQC structure $U_{i}$. We consider two type of surrogate model candidates: graph neural network and random forest, chosen for their ability to capture structural and statistical patterns, respectively. Next, we outline the model architectures and training procedure in detail.
\subsubsection{Graph Neural Network~(GNN)}
GNNs~\cite{gnns} have attracted growing interest as surrogate models for QAS\cite{lukasik2021neural, He2024AMG, he2023gsqas}. 
Designed to process graph-structured data, GNNs learn representations of nodes, edges, or entire graphs by exploiting the relational information encoded in graph topologies. 
Here, we use  a multi-layer Graph Convolutional Network~(GCN)~\cite{kipf2016semi} as surrogate model, which processes quantum circuits represented as Directed Acyclic Graphs (DAGs) to predict circuit performance. 
To enable this, we first transform each PQC into a graph-based data structure that captures the data flow of qubit operations. The network is then trained directly on these graph-based circuit representations. 

Given a quantum circuit defined over a gate set alphabet~$H$, each \textit{node} $V$ in the graph~$G=(V,E)$ corresponds to a quantum gate $g \in H$ incorporated into the circuit. 
The gate type is encoded using one-hot vector representation. 
For parameterized gates, such as $ry(\theta)$, the continuous parameters are appended to the one-hot vector.
Thus, each node in the graph is associated with a \textit{node feature matrix} $feat_{v_i}$ of shape $[|H|, |H|+\underset{g \in H}{max }|\theta_g|]$, where $|H|$ is the total number of gate types in $H$ and $\underset{g \in H}{max }|\theta_g|$ is the maximum number of parameters required by any gate $g \in H$.
This representation enables the GCN to differentiate among various gate types and capture continuous gate parameters when present.

\begin{figure}[tb]
\centerline{\includegraphics[width=0.9\linewidth]{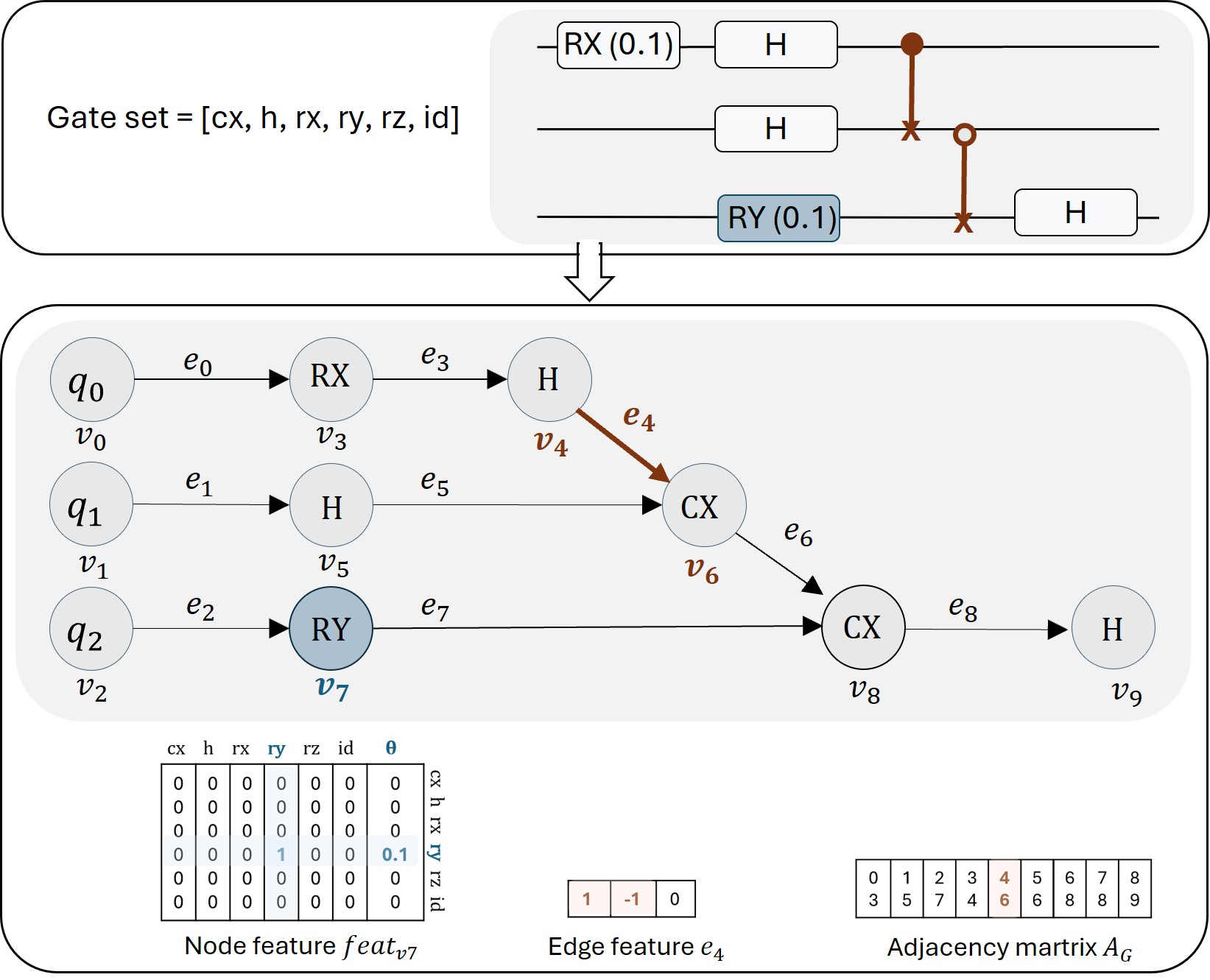}}
\caption{Conversion of a quantum circuit with $n=3$ qubits into a DAG. \textit{Top}: The original quantum circuit along with the gate set alphabet. \textit{Bottom}: The corresponding DAG representation, where each node represents a quantum gate and directed edges capture temporal dependencies along qubit lines. Also shown are (i) the node feature vector for $v_7$, (ii) the edge feature vector for $e_4$ and (iii) the adjacency matrix $A_G$ for the graph.
} 
\label{fig:dag}
\end{figure}

Directed \textit{edges} encode the temporal and causal relationships between gate operations along individual qubits.
A directed edge from the node $v_i$, representing the gate~$g_i$, to the node~$v_j$, representing the gate $g_j$, is added if and only if: both~$g_i$ and~$g_j$ act on the same qubit, and $g_j$
is the next operation on that qubit following $g_i$,~i.e.,~there is no intervening operation on that qubit between $g_i$ and $g_j$. 
If~$v_i$ represents a multi-qubit gate~$g_k$ acting on qubits~$q_c$ and $q_t$, e.g. CNOT, then $v_i$ will have multiple incoming edges: one from the node representing last gate before $g_k$
acting on the qubit~$q_c$ and one from the node representing last gate before $g_k$ on the qubit~$q_t$. 
These edges denote that $g_k$ can only be executed once all those predecessor gates have been executed, capturing the causal structure of the circuit. 
This construction results in linear chains of dependencies along individual qubit timelines. 
When a multi-qubit gate connects two or more such chains, they merge at the corresponding gate node. 
In this way, the graph explicitly encodes both temporal ordering and qubit interaction patterns. 
Each edge~$e_{ij}$ contains an \textit{edge attribute vector}~$feat_{e_{ij}}$ of fixed length $n$, where $n$ corresponds to the number of qubits in the circuit.
Each entry in this vector encodes the involvement of the corresponding qubit in the operation associated with the edge’s destination node. 
Specifically, a value of $1$ indicates that the qubit is the target of a gate (or the only qubit involved in a single-qubit gate); a value of $-1$ indicates that the qubit is the control of a multi-qubit gate, e.g., CNOT or CZ; a value of 0 indicates that the qubit is not involved in the operation.
Finally, the \textit{adjacency matrix} $A_G$ of the graph is represented by a tensor of dimension [$2, |e|$], where $e$ denotes the number of directed edges in the graph. 
Each column of $A_G$ encodes a single edge  $e_{ij} \in E$, with the first row specifying the source node $v_i$ for $e_{ij}$ and the second row specifying the destination node $v_j$. Fig.~\ref{fig:dag} illustrates schematically the described procedure.

\subsubsection{Random Forest} The Random Forest~(RF)~\cite{598994} is a classical machine learning model that constructs an ensemble of decision trees and combines their predictions, typically via averaging, to achieve robust and accurate regression performance. 
To train the RF model that approximates the outcomes of PQCs, we convert the string representation of quantum circuit in OpenQASM into a tensor representation, which serves as input to the RF model. For this purpose, each instruction is processed by encoding the gate type as a one-hot vector, mapping qubit usage to numerical values (+1 for active qubits, –1 for control qubits), and extracting any associated gate parameters. 
The resulting feature vectors are organized into rows, padded to a fixed circuit depth using no-operation entries, and converted into a tensor.

\subsection{Training of the surrogate model} 
\begin{figure}[tb]
\centerline{\includegraphics[width=1\linewidth]{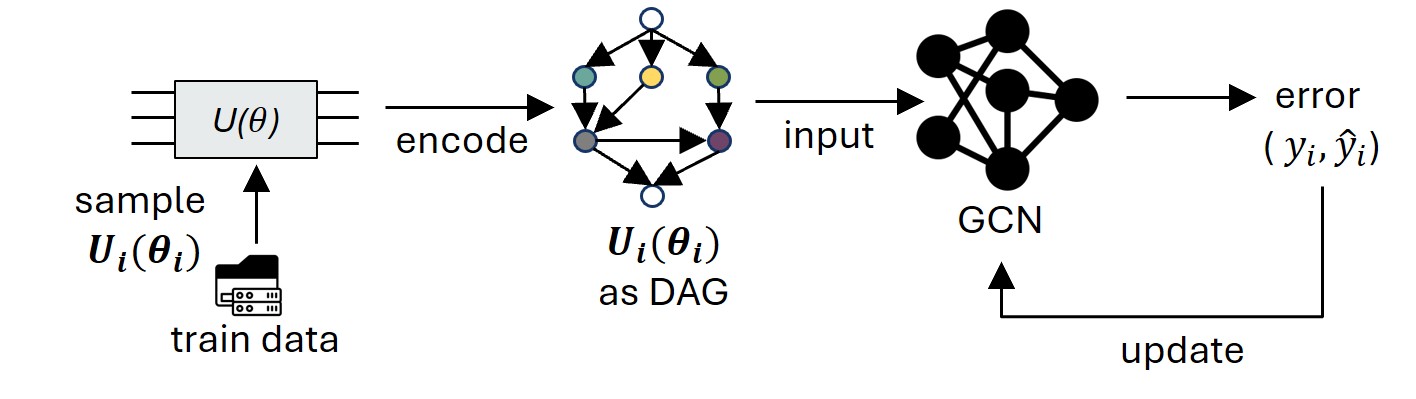}}
\caption{Graph-based training pipeline for the surrogate model}
\label{fig:model_training}
\end{figure}

Surrogate models are trained to approximate the performance metric of quantum circuits by minimizing a Mean Squared Error~(MSE) loss function, as depicted in Fig.~\ref{fig:model_training}.
In addition, we monitor the ranking accuracy of the predictions using Spearman's rank correlation coefficient, which is used in combination with MSE for the model selection during hyperparameter tuning, ensuring that the relative ordering of circuit's performance is accurately captured.

\section{Benchmark Evaluation}\label{sec:evaluation}
{\fontsize{6pt}{7pt}\selectfont
\renewcommand{\arraystretch}{0.9}
\begin{table}[tb]
\caption{Test metrics for surrogate models on~\texttt{GHZ\_a}}

\begin{tabular}{|@{\hskip 0.2pt}c@{\hskip 0.2pt}
                |@{\hskip 0.2pt}c@{\hskip 0.2pt}
                |@{\hskip 0.2pt}c@{\hskip 0.2pt}
                |@{\hskip 0.2pt}c@{\hskip 0.2pt}
                |@{\hskip 0.2pt}c@{\hskip 0.2pt}
                |@{\hskip 0.2pt}c@{\hskip 0.2pt}
                |@{\hskip 0.2pt}c@{\hskip 0.2pt}|}
\hline
\textbf{}                                                                       & \multicolumn{1}{c|}{\textbf{GCN}}    & \multicolumn{1}{c|}{\textbf{\begin{tabular}[c]{@{}c@{}}GCN\\ \_pr\end{tabular}}} & \multicolumn{1}{c|}{\textbf{\begin{tabular}[c]{@{}c@{}}GCN\\ \_aug\end{tabular}}} & \multicolumn{1}{c|}{\textbf{\begin{tabular}[c]{@{}c@{}}GCN\_\\ pr\_aug\end{tabular}}} & \multicolumn{1}{c|}{\textbf{RF}} & \multicolumn{1}{c|}{\textbf{\begin{tabular}[c]{@{}c@{}}RF\\ \_aug\end{tabular}}} \\ \hline
\textbf{\# Samples}                                                             & \multicolumn{6}{c|}{82926}                                                                                                                                                                                                                                                                                                                                                                                                \\ \hline
\textbf{MSE}                                                                    & \multicolumn{1}{c|}{0.0025}          & \multicolumn{1}{c|}{0.0020}                                                      & \multicolumn{1}{c|}{\textbf{0.0018}}                                              & \multicolumn{1}{c|}{0.0023}                                                           & \multicolumn{1}{l|}{0.0151}      & 0.0153                                                                           \\ \hline
\textbf{MAE}                                                                    & \multicolumn{1}{c|}{0.0296}          & \multicolumn{1}{c|}{0.0193}                                                      & \multicolumn{1}{c|}{\textbf{0.0183}}                                              & \multicolumn{1}{c|}{0.0269}                                                           & \multicolumn{1}{l|}{0.0739}      & 0.0791                                                                           \\ \hline
\textbf{RMSE}                                                                   & \multicolumn{1}{c|}{0.0499}          & \multicolumn{1}{c|}{0.0449}                                                      & \multicolumn{1}{c|}{\textbf{0.0425}}                                              & \multicolumn{1}{c|}{0.0480}                                                           & \multicolumn{1}{l|}{0.1229}      & 0.1236                                                                           \\ \hline
\textbf{R\textasciicircum{}2}                                                   & \multicolumn{1}{c|}{0.9246}          & \multicolumn{1}{c|}{0.9389}                                                      & \multicolumn{1}{c|}{\textbf{0.9453}}                                              & \multicolumn{1}{c|}{0.9303}                                                           & \multicolumn{1}{l|}{0.5424}      & 0.7050                                                                           \\ \hline
\textbf{Corr}                                                                   & \multicolumn{1}{c|}{0.9681}          & \multicolumn{1}{c|}{0.9694}                                                      & \multicolumn{1}{c|}{\textbf{0.9724}}                                              & \multicolumn{1}{c|}{0.9693}                                                           & \multicolumn{1}{l|}{0.7669}      & 0.8545                                                                           \\ \hline
\textbf{Spearman}                                                               & \multicolumn{1}{c|}{\textbf{0.8320}} & \multicolumn{1}{c|}{0.8314}                                                      & \multicolumn{1}{c|}{0.8318}                                                       & \multicolumn{1}{c|}{0.8318}                                                           & \multicolumn{1}{l|}{0.7337}      & 0.7870                                                                           \\ \hline
\textbf{\begin{tabular}[c]{@{}c@{}}Accuracy\\ ($|err| \leq= 0.1$)\end{tabular}} & \multicolumn{1}{c|}{0.9584}          & \multicolumn{1}{c|}{0.9601}                                                      & \multicolumn{1}{c|}{\textbf{0.9636}}                                              & \multicolumn{1}{c|}{0.9595}                                                           & \multicolumn{1}{l|}{0.7320}      & 0.7042                                                                           \\ \hline
\end{tabular}
\label{tab:gs1_evaluation}
\end{table}
}

{\fontsize{6pt}{7pt}\selectfont
\renewcommand{\arraystretch}{0.9}
\begin{table}[tb]
\caption{Test metrics for surrogate models on~\texttt{GHZ\_b}.}

\begin{tabular}{|@{\hskip 0.1pt}c@{\hskip 0.1pt}
                |@{\hskip 0.1pt}c@{\hskip 0.1pt}
                |@{\hskip 0.1pt}c@{\hskip 0.1pt}
                |@{\hskip 0.1pt}c@{\hskip 0.1pt}
                |@{\hskip 0.1pt}c@{\hskip 0.1pt}
                |@{\hskip 0.1pt}c@{\hskip 0.1pt}
                |@{\hskip 0.1pt}c@{\hskip 0.1pt}|}
\hline
\textbf{}                                                                       & \multicolumn{1}{c|}{\textbf{GCN}} & \multicolumn{1}{c|}{\textbf{\begin{tabular}[c]{@{}c@{}}GCN\\ \_pr\end{tabular}}} & \multicolumn{1}{c|}{\textbf{\begin{tabular}[c]{@{}c@{}}GCN\\ \_aug\end{tabular}}} & \multicolumn{1}{c|}{\textbf{\begin{tabular}[c]{@{}c@{}}GCN\_\\ pr\_aug\end{tabular}}} & \multicolumn{1}{c|}{\textbf{RF}} & \multicolumn{1}{c|}{\textbf{\begin{tabular}[c]{@{}c@{}}RF\\ \_aug\end{tabular}}} \\ \hline
\textbf{\begin{tabular}[c]{@{}c@{}}\# \\ Samples\end{tabular}}                  & \multicolumn{6}{c|}{97753}                                                                                                                                                                                                                                                                                                                                                                                             \\ \hline
\textbf{MSE}                                                                    & \multicolumn{1}{c|}{0.0072}       & \multicolumn{1}{c|}{0.0074}                                                      & \multicolumn{1}{c|}{\textbf{0.0069}}                                              & \multicolumn{1}{c|}{\textbf{0.0069}}                                                  & \multicolumn{1}{l|}{0.0184}      & 0.0187                                                                           \\ \hline
\textbf{MAE}                                                                    & \multicolumn{1}{c|}{0.0467}       & \multicolumn{1}{c|}{0.0491}                                                      & \multicolumn{1}{c|}{\textbf{0.0451}}                                              & \multicolumn{1}{c|}{0.0457}                                                           & \multicolumn{1}{l|}{0.1044}      & 0.1054                                                                           \\ \hline
\textbf{RMSE}                                                                   & \multicolumn{1}{c|}{0.0847}       & \multicolumn{1}{c|}{0.0863}                                                      & \multicolumn{1}{c|}{\textbf{0.0831}}                                              & \multicolumn{1}{c|}{\textbf{0.0831}}                                                  & \multicolumn{1}{l|}{0.1357}      & 0.1366                                                                           \\ \hline
\textbf{R\textasciicircum{}2}                                                   & \multicolumn{1}{c|}{0.8352}       & \multicolumn{1}{c|}{0.8292}                                                      & \multicolumn{1}{c|}{\textbf{0.8413}}                                              & \multicolumn{1}{c|}{0.8415}                                                           & \multicolumn{1}{l|}{0.5770}      & 0.5715                                                                           \\ \hline
\textbf{Corr}                                                                   & \multicolumn{1}{c|}{0.9140}       & \multicolumn{1}{c|}{0.9109}                                                      & \multicolumn{1}{c|}{0.9175}                                                       & \multicolumn{1}{c|}{\textbf{0.9178}}                                                  & \multicolumn{1}{l|}{0.7838}      & 0.7811                                                                           \\ \hline
\textbf{Spearman}                                                               & \multicolumn{1}{c|}{0.8743}       & \multicolumn{1}{c|}{0.8717}                                                      & \multicolumn{1}{c|}{0.8758}                                                       & \multicolumn{1}{c|}{\textbf{0.8776}}                                                  & \multicolumn{1}{l|}{0.7718}      & 0.7692                                                                           \\ \hline
\textbf{\begin{tabular}[c]{@{}c@{}}Accuracy\\ ($|err| \leq= 0.1)$\end{tabular}} & \multicolumn{1}{c|}{0.8566}       & \multicolumn{1}{c|}{0.8474}                                                      & \multicolumn{1}{c|}{0.8661}                                                       & \multicolumn{1}{c|}{\textbf{0.8635}}                                                  & \multicolumn{1}{l|}{0.5663}      & 0.5611                                                                           \\ \hline
\end{tabular}
\label{tab:gs2_evaluation}
\end{table}
}

In this section, we perform an empirical analysis to demonstrate the application of the proposed benchmark and gain insight into the performance of surrogate models.
As outlined in previous sections, we train two types of models, GCN and RF, on two tasks: state preparation and classification. To implement the GCN architecture, we use \textit{PyTorch Geometrics}~\cite{Fey/Lenssen/2019}, while the RF model is implemented using~\textit{sklearn}~\cite{pedregosa2011scikit}. 

For the RL agent, we employ the Proximal Policy Optimization~(PPO)~\cite{schulman2017proximal} and Deep Q Network~(DQN)~\cite{mnih2013playing} algorithms, using the implementations provided by~\textit{Stable-Baselines3}~\cite{stable-baselines3}. 
For both quantum state preparation and classification tasks, the reward function is designed to guide the agent towards learning an optimal strategy $\pi^*$ to sequentially generate a PQC with high performance and low depth. 
At each time step $t$, the agent observes the current environment state $s_t$, represented by the full quantum state vector. Based on this observation, it selects a discrete action $a_t$, corresponding to either a single quantum gate or an entire layer of gates.

The implementation of the GA relies on the \textit{DEAP} library~\cite{DEAP_JMLR2012}. 
The GA optimizes the generation of PQCs by mimicking natural selection. Starting from a randomly generated population, new candidates are produced through mutations, i.e., random insertions or deletions of gates or layers. Each circuit is evaluated using a fitness function, such as fidelity or classification accuracy, and the individuals with the highest fitness values are selected for the next generation. 
This process gradually refines the population towards more optimal solutions.

\subsection{State Preparation} As described in Section~\ref{sec:methods}, we consider two search spaces for the state preparation task: \texttt{GHZ\_a} and \texttt{GHZ\_b}. For each search space, the data is available in two variants, non-augmented (\texttt{gs1\_base}, \texttt{gs2\_base}) and augmented~(\texttt{gs1\_aug}, \texttt{gs2\_aug}).
For each data variant, we train and tune both a GCN and a RF models. We refer to models trained on augmented data as \textit{GCN\_aug} and \textit{RF\_aug}, and to those trained on the non-augmented data as \textit{GCN} and \textit{RF}.
In addition, we evaluate an extended version of GCN model by providing an auxiliary input: a lightweight, train-free proxy designed to indicate whether a given circuit has the structural potential to produce a maximally entangled state.
This proxy analyzes the structure of a circuit and assigns a numerical score of~$1.0$ if both (i) superposition is possible and (ii) a sufficient entanglement structure is present; otherwise, it returns~$0.0$. The resulting proxy score is appended to the original model input as an additional feature.
Models trained with this extended input are referred to as \textit{GCN\_pr} and \textit{GCN\_pr\_aug}.
Details on all model architectures and hyperparameters are available in our repository.

Tables~\ref{tab:gs1_evaluation} and~\ref{tab:gs2_evaluation} summarize the performance metrics of the trained models, evaluated on test sets sampled from the original (non-augmented) data in the \texttt{GHZ\_a} and \texttt{GHZ\_b} search spaces, respectively. 
In both search spaces, the GCN models outperform the RF models. 
Among the GCN models, performance is comparable, with models trained on augmented data showing a slight improvement across all metrics, highlighting the beneficial effect of data augmentation. 
The integration of the train-free proxy input in GCN models leads to minor improvements in some cases. 
However, this effect is not uniform across all datasets, suggesting that further investigation is needed to understand its impact more thoroughly.

\begin{figure}[tb]
\centerline{\includegraphics[width=1\linewidth]{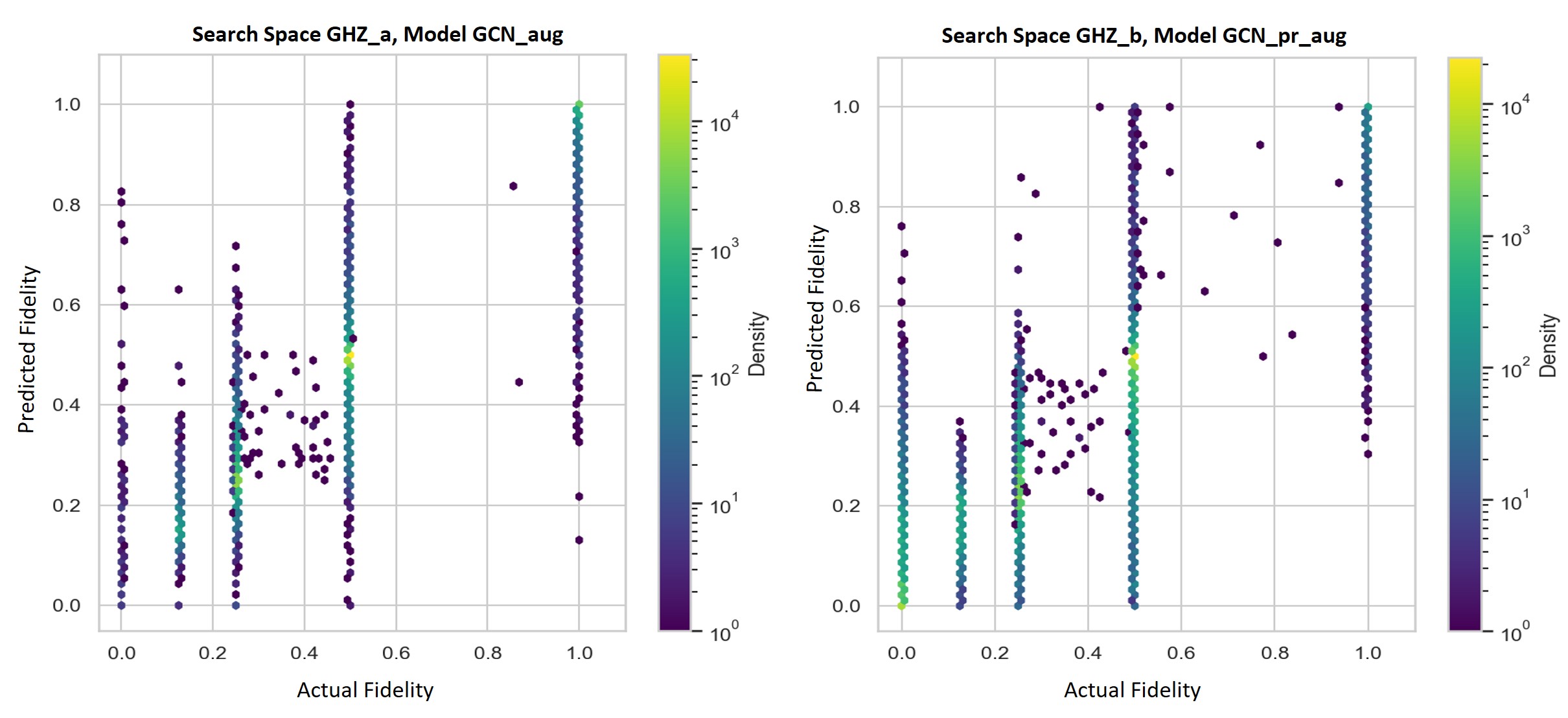}}
\caption{Predicted vs. actual fidelity for surrogate models on the state preparation. Left: Search space \texttt{GHZ\_a}, GCN trained on augmented data (GCN\_aug). Right: \texttt{GHZ\_b}, GCN with proxy, trained on augmented data (GCN\_pr\_aug).}
\label{fig:surr_evaluation_ghz}
\end{figure}

\begin{figure}[t]
\centerline{\includegraphics[width=1\linewidth]{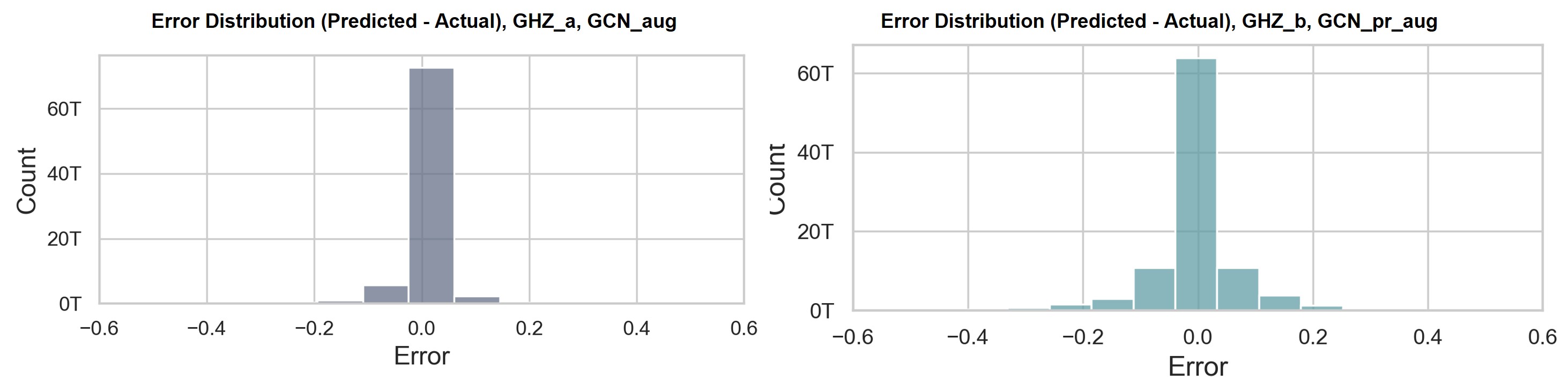}}
\caption{Error distribution calculated on test subsets. Top: \textit{GCN\_aug} on the \texttt{GHZ\_a} test subset. Bottom: \textit{GCN\_pr\_aug} on the \texttt{GHZ\_b} test subset.}
\label{fig:surr_evaluation_errors_ghz}
\end{figure}

Next, we further evaluate the best performing models from each search space, i.e.,~\textit{GCN\_aug} for the \texttt{GHZ\_a} and \textit{GCN\_pr\_aug} for \texttt{GHZ\_b}, by comparing the distributions of their predicted outputs to the ground truth circuit fidelity, depicted at Fig.~\ref{fig:surr_evaluation_ghz}. 
Fig.~\ref{fig:surr_evaluation_errors_ghz} outlines error distributions of the models.
Although the GCN model achieves high accuracy on \texttt{GHZ\_a}, two key problem areas are evident from the plots. 
First, the model shows difficulties in correctly identifying circuits with a true fidelity of 0.0, although this issue is relatively minor. 
More concerning is the model performance on high-fidelity circuits. While most are predicted accurately, a subset of circuits is erroneously classified as having a fidelity $\leq 0.8$.
At the same time, some circuits with a true fidelity around 0.5 are misclassified, with predictions in the range of 0.6 to 0.8, or even higher than 0.8 in a few cases.
These overlaps are problematic, as they could lead to incorrect predictions and inconsistent feedback for QAS methods. 
This could negatively impact the QAS performance, particularly when high-fidelity circuits are sparse, as is the case for~\texttt{GHZ\_b}, where the overlap is even more visible.
This issue is addressed in the experiments through a thresholding strategy.

\begin{figure*}[tb]
\centerline{\includegraphics[width=1\textwidth]{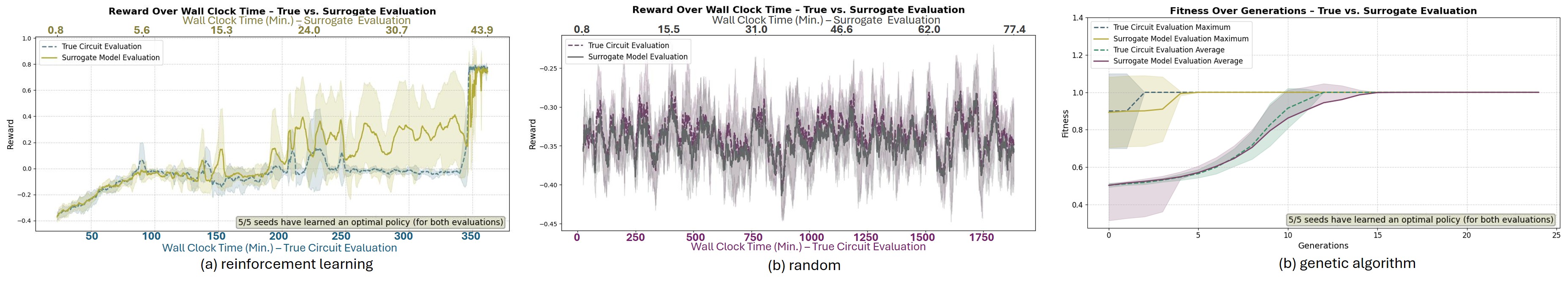}}
\caption{Performance of QAS methods for the \texttt{GHZ\_a} search space on the real circuit evaluation and evaluation using the \texttt{GCN\_aug} model. All evaluations are conducted with 5 independent seeds. \textit{Left}: Wall clock time / learning curve for the RL agent, DQN. \textit{Middle}: Wall clock time / learning curve for the random approach. \textit{Right}: Mean and average fitness per generation for the GA. For RL and random, a smoothing window of 50 is applied. For all 5 seeds, the surrogate model successfully guides the QAS method in learning RL an optimal policy and GA identifying the best individual.}
\label{fig:gs1_evaluation}
\end{figure*}
To evaluate the performance of the surrogate models within the context of QAS and to demonstrate the effectiveness of the proposed benchmark, we compare the learning curves of various QAS methods under two distinct circuit evaluation procedures. 
In the first setting, QAS is executed using RL, GA, and random search, each with five fixed seeds. Candidate PQCs are evaluated based on their training, execution on a simulator, and comparison of the resulting statevectors to the target statevector.
In the second setting, the same methods are applied, but the circuits are assessed using the surrogate model. 
For each search space, we select the best performing model and choose seeds and algorithmic setting, which were not involved in the data collection process, e.g,~DQN policy instead of PPO for~RL. All experiments are conducted on CPU.

Fig.~\ref{fig:gs1_evaluation} presents the results for \texttt{GHZ\_a}. 
The trainable methods, RL and GA, successfully converged to an optimal policy or best individual in all five runs, both under true evaluation and when using the surrogate model.
The learning curves for true and surrogate-based evaluations show minor discrepancies, with the surrogate model still capturing the overall trend accurately.
In the case of the random search, the surrogate-based evaluation mirrors the true evaluation trend. 
A notable advantage of using the proposed surrogate benchmark is the significant time speedup it offers. 
For instance, in the case of RL, the true evaluation requires approximately 6 hours, whereas the surrogate-based evaluation completes in about 45 minutes. The speedup is even more visible for random search, with runtimes of approximately 31 hours for true evaluation compared to just 1.2 hours using the proposed benchmark.
\begin{figure}[tb]
\centerline{\includegraphics[width=1\linewidth]{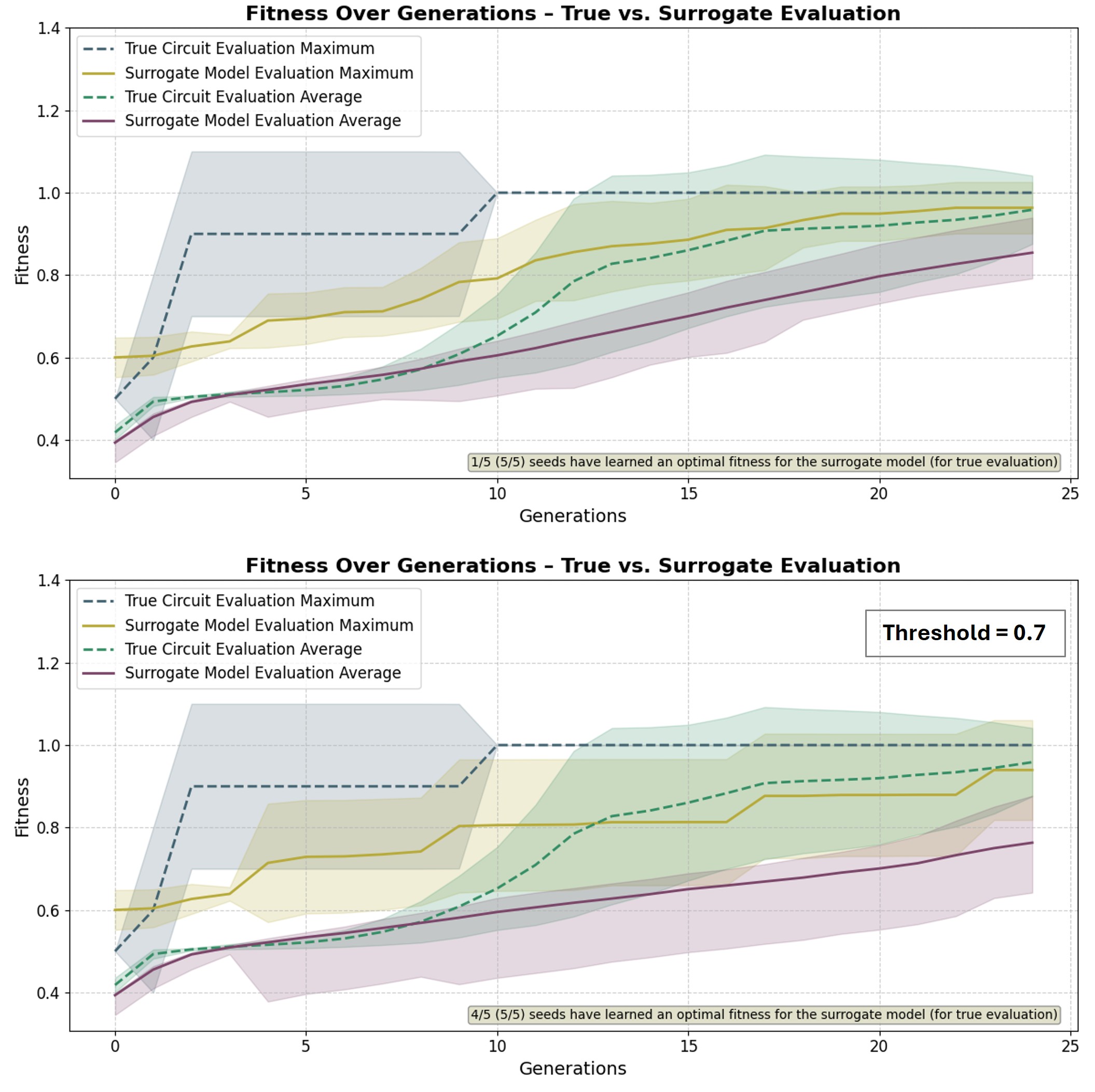}}
\caption{Learning curve of GA on \texttt{GHZ\_b} with 5 independent seeds. \textit{Top}: Comparison of real circuit evaluation versus evaluation using the surrogate model. \textit{Bottom}: Comparison of real circuit evaluation versus surrogate model evaluation with a threshold of $0.7$; if the predicted fidelity is $\geq 0.7$, real evaluation is performed; otherwise, the surrogate model is used. For the threshold-based approach, GA with the surrogate model identified the best individuals in 3 out of 5 seeds, as compared to those found through QAS with real evaluation.}
\label{fig:surr_evaluation_gs2}
\end{figure}

Fig.~\ref{fig:surr_evaluation_gs2} outlines the execution of GA on the search space~\texttt{GHZ\_b}. When QAS is executed using surrogate evaluation only, one out of five seeds converges to an individual with fidelity 1.0. 
In contrast, QAS with real evaluation identifies high-performing circuits in all five seeds. However, using a threshold of $0.7$, i.e., only moving on to the actual evaluation when the predicted fidelity exceeds $0.7$, leads to high fidelity individuals being found in four out of five seeds.

\subsection{Classification}
\begin{figure}[tb]
\centerline{\includegraphics[width=1\linewidth]{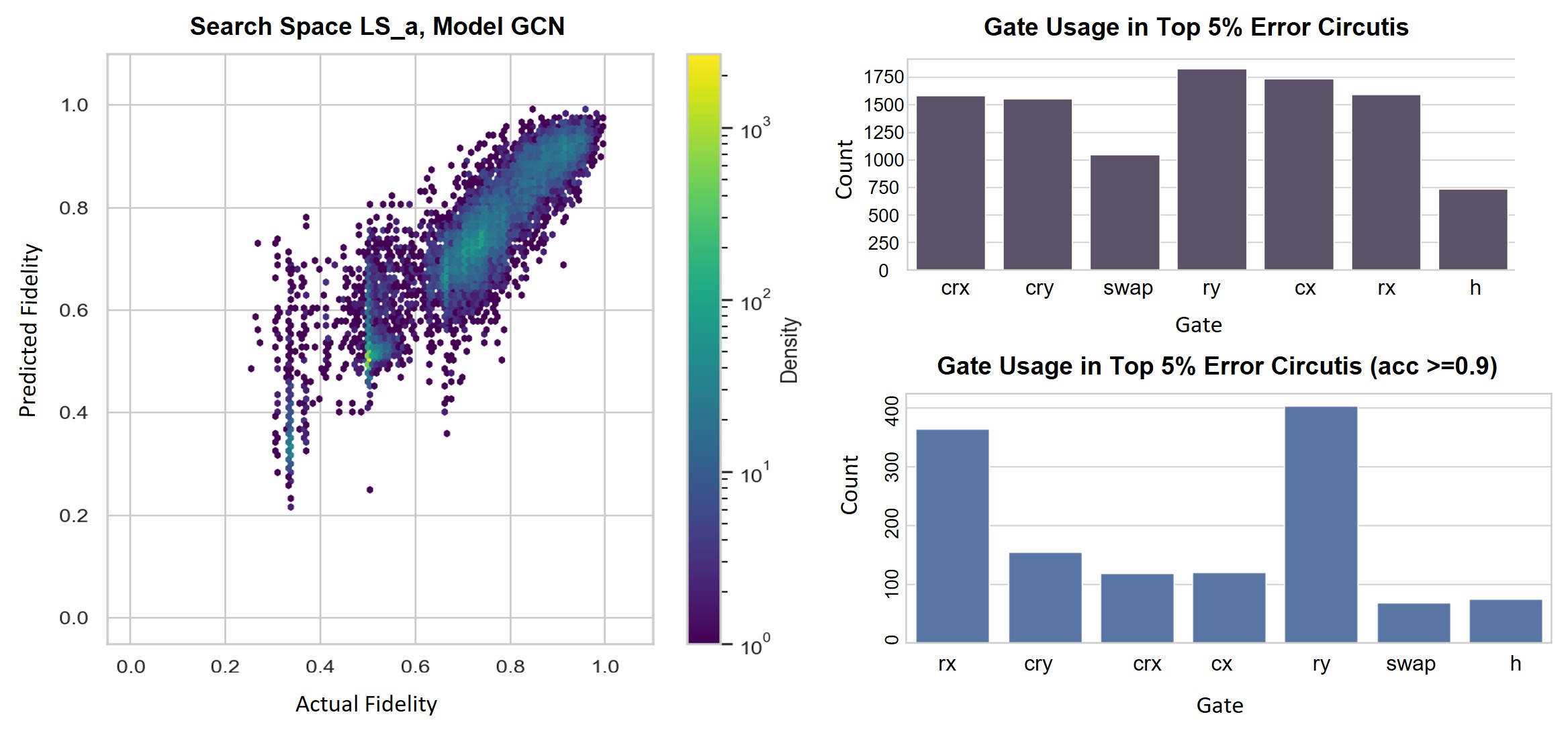}}
\caption{Performance evaluation of the GCN model on \texttt{LS\_a}.
\textit{Left}: Predicted vs. actual training accuracy of PQCs, evaluated on the test subset.
\textit{Right}: Gate usage statistics for circuits with prediction errors $\geq 0.1$ (top), and for high-performing circuits (training accuracy $\geq 0.9$) with prediction errors $\geq 0.1$ (bottom).} 
\label{fig:ml_eval}
\end{figure}

\begin{figure}[tb]
\centerline{\includegraphics[width=1\linewidth]{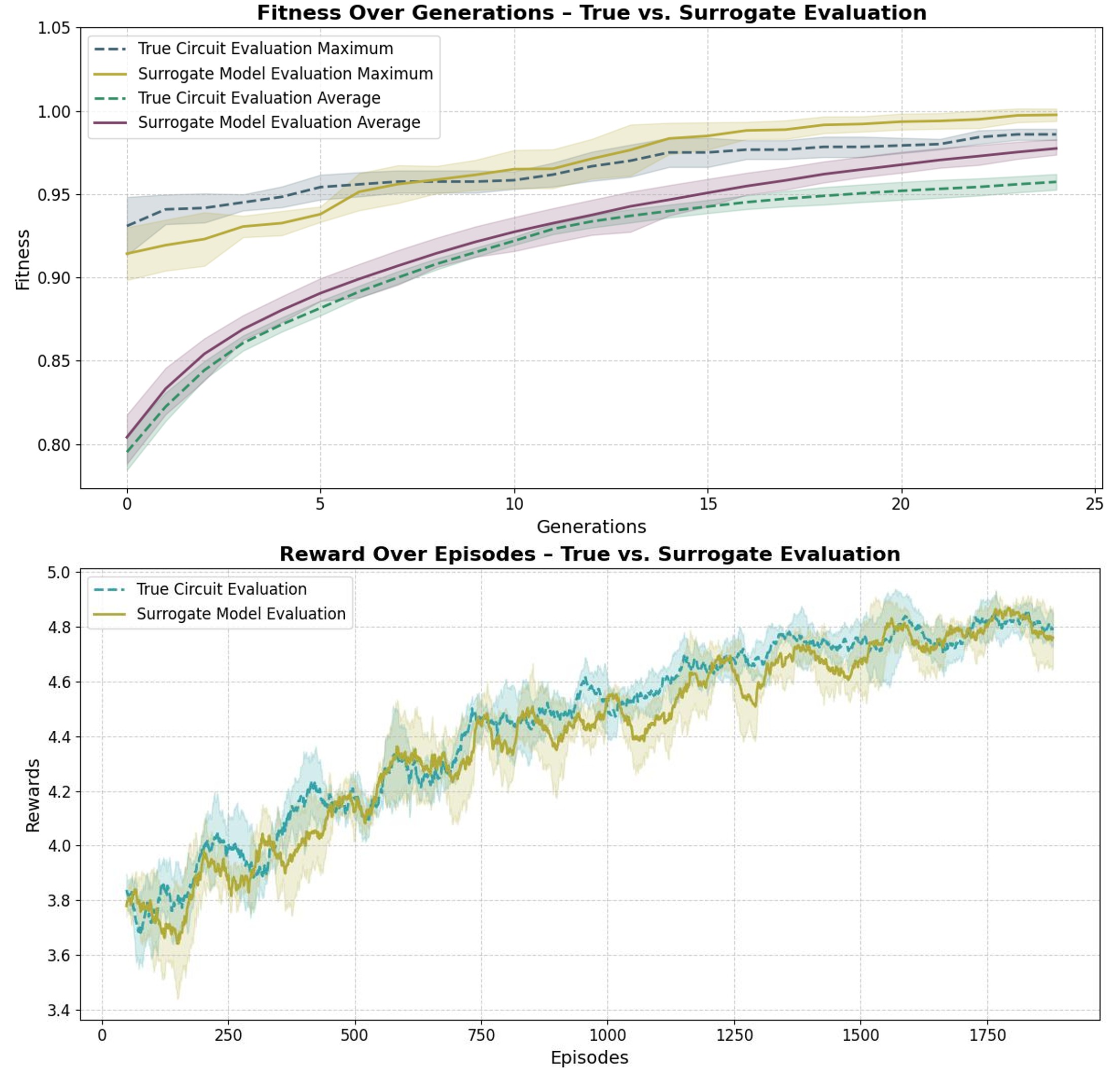}}
\caption{Learning curves of QAS methods on \texttt{LS\_a}. \textit{Top}: GA performance using true versus surrogate-based evaluations, averaged over 5 independent runs. \textit{Bottom}: RL performance with true and surrogate evaluations, shown across 5 seed histories recorded during the data generation process.}
\label{fig:ml_eval_ga_rl}
\end{figure}

For the data classification task, we similarly train a GCN and a RF model. Here, only one search space~\texttt{LS\_a} and one dataset~\textit{ls\_base} exist. Table~\ref{tab:ml_evaluation} shows the performance metrics of the trained models on a test subset. 
\begin{table}[tb]
\caption{Test metrics for surrogate models for the \texttt{LS\_a.}}
\begin{center}
\begin{tabular}{|c|cc|}
\hline
\textbf{}                     & \multicolumn{1}{c|}{\textbf{GCN}}    & \textbf{RF} \\ \hline
\textbf{\# Samples}           & \multicolumn{2}{c|}{34569}                         \\ \hline
\textbf{MSE}                  & \multicolumn{1}{c|}{\textbf{0.0026}} & 0.0059      \\ \hline
\textbf{MAE}                  & \multicolumn{1}{c|}{\textbf{0.0315}} & 0.0564      \\ \hline
\textbf{RMSE}                 & \multicolumn{1}{c|}{\textbf{0.0507}} & 0.0767      \\ \hline
\textbf{R\textasciicircum{}2} & \multicolumn{1}{c|}{\textbf{0.8949}} & 0.7571      \\ \hline
\textbf{Corr}                 & \multicolumn{1}{c|}{\textbf{0.9480}} & 0.8728      \\ \hline
\textbf{Spearman}             & \multicolumn{1}{c|}{\textbf{0.9298}} & 0.8742      \\ \hline
\textbf{\begin{tabular}[c]{@{}c@{}}Accuracy\\ ($|err| \leq= 0.1)$\end{tabular}}             & \multicolumn{1}{c|}{\textbf{0.9412}} & 0.830       \\ \hline
\end{tabular}
\end{center}
\label{tab:ml_evaluation}
\end{table}
Similarly to the prior search spaces, for~\texttt{LS\_a} the GCN outperforms RF across all metrics. 
The density of predictions in Fig.~\ref{fig:ml_eval} indicates a diagonal alignment, while two clusters emerge around training accuracies of approximately 0.5 and 0.3. 
An analysis of gate usage in PQCs with prediction errors greater than~0.1 reveals no specific gate types that systematically degrade prediction quality, suggesting that the model is unbiased with respect to gate composition.
Figure~\ref{fig:ml_eval_ga_rl} shows the learning curves of the GA and RL methods, averaged over multiple runs. The surrogate-based evaluation closely tracks the true evaluation and effectively supports the search process. However, our experiments revealed a clear drop in evaluation accuracy when the surrogate was required to distinguish between high-performing circuits. In particular, once circuit accuracy exceeds 95\%, the model often fails to produce predictions close to the true values. Yet, in this high-accuracy regime, prediction precision becomes critical for convergence to the optimal solution, representing a notable challenge. 
To mitigate this, we recommend applying a confidence threshold of 95\% when convergence to an optimal solution is required.

\section{Discussion}\label{sec:limitations}
In this work, we present, to the best of our knowledge, the first initiative to establish a surrogate benchmark for QAS. 
With this contribution, we aim to support the development of reproducible evaluation practices in the field and to encourage the automated quantum machine learning (AutoQML) community to devote greater attention to benchmark-driven research.
Several promising directions remain for future research to further expand and refine this approach. 

A key direction involves a deeper investigation into the behavior and limitations of surrogate models. 
Conducting ablation studies to evaluate the influence of individual circuit parameters and specific gate types on predictive performance could provide valuable insights. Additionally, it remains an open question how much data is required to effectively fine-tune the surrogate model for new or slightly shifted search spaces.
We also observe that model performance is not consistent across all spaces. 
For example, the~\texttt{GHZ\_b} search space presents notable challenges for surrogate models.
This observation is consistent with findings from~\cite{nakayama2023vqe}, which highlight the difficulty classical models, particularly GNNs, face when attempting to classify or approximate the behavior of quantum circuits. Such challenges may partially account for the reduced effectiveness of surrogates in this domain.
To address these limitations, future work could explore the use of quantum models as surrogates. 
These models may be better suited to capture quantum-specific correlations and could leverage richer input representations that integrate both circuit topology and properties of the quantum states they generate.
Furthermore, alternative learning paradigms such as classification or clustering rather than regression may prove more effective for certain tasks, e.g. the state preparation.

When working with limited, imbalanced, or noisy datasets, data augmentation is essential to improve the effectiveness and robustness of classical model training. 
In this work, we employed basic augmentation technique by recompiling circuits using subsets of the original gate sets. However, more advanced augmentation strategies can be used in future work, e.g., recently introduced tool KetGPT~\cite{apak2024ketgpt}, which enables circuit augmentation using transformer-based models. 

While SQuASH currently focuses on relatively simple tasks, already enabling substantial speedups in experimentation and benchmarking, the scope of tasks and search spaces should be expanded toward more complex scenarios, such as PQCs optimized using stochastic methods or noise-aware setups. 
Moreover, applying the benchmark to open-source implementations from the broader QAS and AutoQML community would enable more comprehensive validation and foster comparability across studies.

QAS methods that aim to jointly optimize both circuit structure and parameters, such as \texttt{qcd-gym}\cite{altmann2023challenges} or approaches based on diffusion models\cite{furrutter2024quantum, nakaji2024generative}, are currently not directly supported by the surrogate evaluation, as the surrogate models assume input circuits with parameters initialized in the range 
[-0.01; 0.01] and subsequently trained. 
Nevertheless, these methods may still benefit from the defined search spaces for performance comparison and from the training data for the surrogate models, e.g., to identify effective PQC patterns or train generative models.

\section{Conclusion and Outlook}\label{sec:conclusion}
In this work, we introduced SQuASH, a benchmark designed to accelerate and unify the evaluation of QAS methods.
By leveraging surrogate models trained on a curated dataset of PQCs, SQuASH significantly reduces the computational cost typically associated with QAS, while guiding the search process towards optimal solutions.
Our benchmark provides multiple pretrained surrogate models, supports various QAS strategies and offers an open-source implementation that can be easily integrated into custom workflows and extended for future research.
While it currently focuses on relatively simple search spaces, our results highlight the potential of surrogate-based QAS benchmarks to speed up experimentation and enable broader method comparisons. 

\bibliographystyle{IEEEtran}
\bibliography{IEEEabrv, main}

\end{document}